# Hybridized Feature Extraction and Acoustic Modelling Approach for Dysarthric Speech Recognition

Megha Rughani
Electronics & Communication
Marwadi Education Foundation's Group of Institution
Rajkot, India
megharughani@gmail.com

D. Shivakrishna
Electronics & Communication
Marwadi Education Foundation's Group of Institution
Rajkot, India
d.shivakrishna@marwadieducation.edu.in

*Abstract*—Dysarthria is malfunctioning of motor speech caused by faintness in the human nervous system. It is characterized by the slurred speech along with physical impairment which restricts their communication and creates the lack of confidence and affects the lifestyle. This paper attempt to increase the efficiency of Automatic Speech Recognition (ASR) system for unimpaired speech signal. It describes state of art of research into improving ASR for speakers with dysarthria by means of incorporated knowledge of their speech production. Hybridized approach for feature extraction and acoustic modelling technique along with evolutionary algorithm is proposed for increasing the efficiency of the overall system. Here number of feature vectors are varied and tested the system performance. It is observed that system performance is boosted by genetic algorithm. System with 16 acoustic features optimized with genetic algorithm has obtained highest recognition rate of 98.28% with training time of 5:30:17.

## I. Introduction

Speech is very essential part of our lives. It is one of the most prominent form of senses through which human easily communicate. It is the speech which assists one develop and nature the relation among the society. The consequences and nature of speech is very less cared while speaking, but in real it is very complex task. It is the combined process of hearing and speaking generated by motor muscles co-ordinated by the brain muscles. However, it is misfortune that some lacks integration of all the above required processes and they are said to have communication disorder. If communication does not occur as it should, the disruption is in the process, not the individual. There are many reasons for the disruption in communication like hearing loss, speaking disability, or lack of co-ordination of neural muscles. Dysarthria is one such impairment which result in slurred speech.

Dysarthria is caused due to reduced control of neuro-motor muscles. It results into slurred speech as articulation is mainly affected. Insertion, deletion and repetition of phoneme reduce the intelligibility of speech signal. Severity of dysarthric speech affects the intelligibility of speech. It is caused due to brain tumour, celebral palsy, Parkinson diseases, head injury and many more. It lessens the controlling portion of brain which is involved in planning, execution and controlling of the specific affected organ along with motor speech disorder. Lungs, larynx, vocal tract movement, lip movement are basically affected. Mostly they are handicapped. [1, 2]

Various clinical treatments including exercise of motor muscles were carried out to increase the strength in order to improve articulation, phonation, and resonance. Special therapy like principles of motor learning are carried out by speech language pathologist but these are very time consuming and tedious to be followed. Assistive technology like Automatic Speech Recognition (ASR) helps in recognition and synthesis of unintelligible speech into intelligible form. [3, 4]

State-of-art of ASR system implemented with normal speakers for automation. But Dysarthric speech is different due to difference in articulation so it gives less recognition when trained on simple ASR system. So, it is required to develop a system specifically for Dysarthric patient.

Spectral and cepstral features are extracted from the input raw speech data which are modelled by various classifiers using acoustic modelling technique and looked up into dictionary to find similar match and accordingly generates the text output. Here Log RASTA Perceptive Linear Prediction (Log RASTA PLP) hybridized with Artificial Neural Network (ANN) is used for feature extraction which are acoustically modelled using hybrid Hidden Markov Model (HMM) /Artificial Neural Network (ANN) technique. Genetic algorithm (GA) is applied for the optimization of the parameters and system performance is compared without and with GA.

## II. PREVIOUS WORK

Here in this section state of art of methods and combination of different technologies developed is presented for Dysarthric as well as normal dataset. For Dysarthric speech recognition rate is not achieved up to the desired level because of the difficulty mentioned in previous section.

Harsh Sharma & Mark Johnson [5] has applied various different algorithms and its combination for the adaptation of the system based on HMM acoustic modelling technique like Maximum A-posterior Probability (MAP) adaptation for Speaker independent (SI) system and Transition probability matrix (Linear interpolation) for Speaker Dependent (SD) and Speaker Adapted (SA) with 12 PLP coefficients along with velocity and acceleration coefficients forming 39 dimensional acoustic feature vector using UA speech database and concluded that adaptation of various parameters leads to increase in word recognition rate or overall system.

Harsh Sharma & Mark Johnson [6] have attempted to explicit modelling by first considering the mismatch between unimpaired and Dysarthric speech among the population and model is prepared. In the second stage this model acts as initial model for the adaptation. Background Interpolated (BI) and MAP adaptation are used for the HMM based system using UA Speech database and PLP feature extraction technique and achieved 4.16% -82.07% of WRA.

Santiago & Caballero [7] found out the best combination of HMM parameters like its topology, number of states and Gaussian mixture components using evolutionary algorithm like Genetic Algorithm. For Speaker dependent approach Bakis topology with 7 states having 13 Gaussian mixture components perform well for some speakers but it cannot be generalized for all speaker as phoneme characteristics varies widely. WRA of 47.27% - 81.22% is obtained by using GA optimized HMM system having Bakis topology.

Santiago & Caballero [8] performed integration of different pronunciation pattern by weighting the responses of an Automatic Speech Recognition (ASR) system when different language model restrictions are set. They performed confusion matrix modelling with weighted Finite State Transducer (WFST) implemented with extended Metamodels (MM) along with evolutionary algorithm like Genetic Algorithm. Comparison of baseline HMM, baseline MM, micro GA and MM built with improved GA was carried and MM with built in GA outperforms among all with WRA of 42.6% - 77.54% which is comparatively similar to previous one implementing HMM and GA algorithm.

Shahamiri & Salim [9] have tried to find out best performing set of MFCC feature extracted set for the usage of ANN acoustic modelling technique and stated that SD and SA have poor performance in terms of speaker's generalizability, so adopted SI approach in his work. Feed forward back propagation training algorithm was implemented using UA speech database and found out that Mel Cepstrum with 12 coefficients, each utterance represented by 264 vector features outperforms well.

Joel Pinto & Hermansky [10] analyze a simple hierarchical architecture consisting of two multilayer perceptron (MLP) classifiers. The first MLP classifier is trained using standard acoustic features. The second MLP is trained using the posterior probabilities of phonemes estimated by the first, but with a long temporal context of around 150-230ms. Here 3 layer MLP architecture is implemented in which temporal context of 90ms is taken of acoustic features obtained using MFCC or PLP technique. These become input to MLP layer 1; again temporal context of 150-230ms is carried and applied to MLP layer 2. On an average recognition rate of 71.6% and 63.3% for TIMIT and CTS database is obtained.

Lilia Lazli & Mounir [11] compared two different approach using speech and biomedical database which are: 1) Multi-Network Radial Basis Function (RBF) / Learning Vector Quantization (LVQ) structure, 2) Hybrid HMM/MLP approach along with K-means clustering algorithm for normal speech and obtained on an average of 90% WRA for HMM/MLP hybrid approach which performs better than multi-network RBF/LVQ method.

Lilia Lazli & Mounir [12] compared five different methods which are: (1) Multi network RBF/LVQ structure (2) Discrete Hidden Markov Models (HMM) (3) Hybrid HMM & MLP system using a Multi-Layer Perceptron (MLP) to estimate the HMM emission probabilities and using the K-means algorithm for pattern clustering (4) Hybrid HMM-MLP system using the Fuzzy C-Means (FCM) algorithm for fuzzy pattern clustering and (5) Hybrid HMM-MLP system using the Genetic Algorithm using three different database (unimpaired speech) along with biomedical speech database. Hybrid HMM/MLP along with GA using Log RASTA PLP feature extraction obtained on an average of 93.5% and this technique outperforms in comparison of other considered technique.

It is clearly concluded from literature review that hybridized approach the recognition of Dysarthric speech has not been practiced. Here it is an attempt to increase the recognition rate by hybridized feature extraction and acoustic modelling technique. Use of ANN for clustering improves the efficiency of the system. Study explains that hybridized acoustic modelling consisting of ANN and HMM can used benefits of both the system and evolutionary algorithm can optimize one of the HMM parameter in order to improve recognition rate.

## III. METHODS & MATERIALS

This section describes system implementation and database used. It gives detailed description of system adopted with neat block diagram. Database which is used in the research work is UA Speech database which is described briefly.

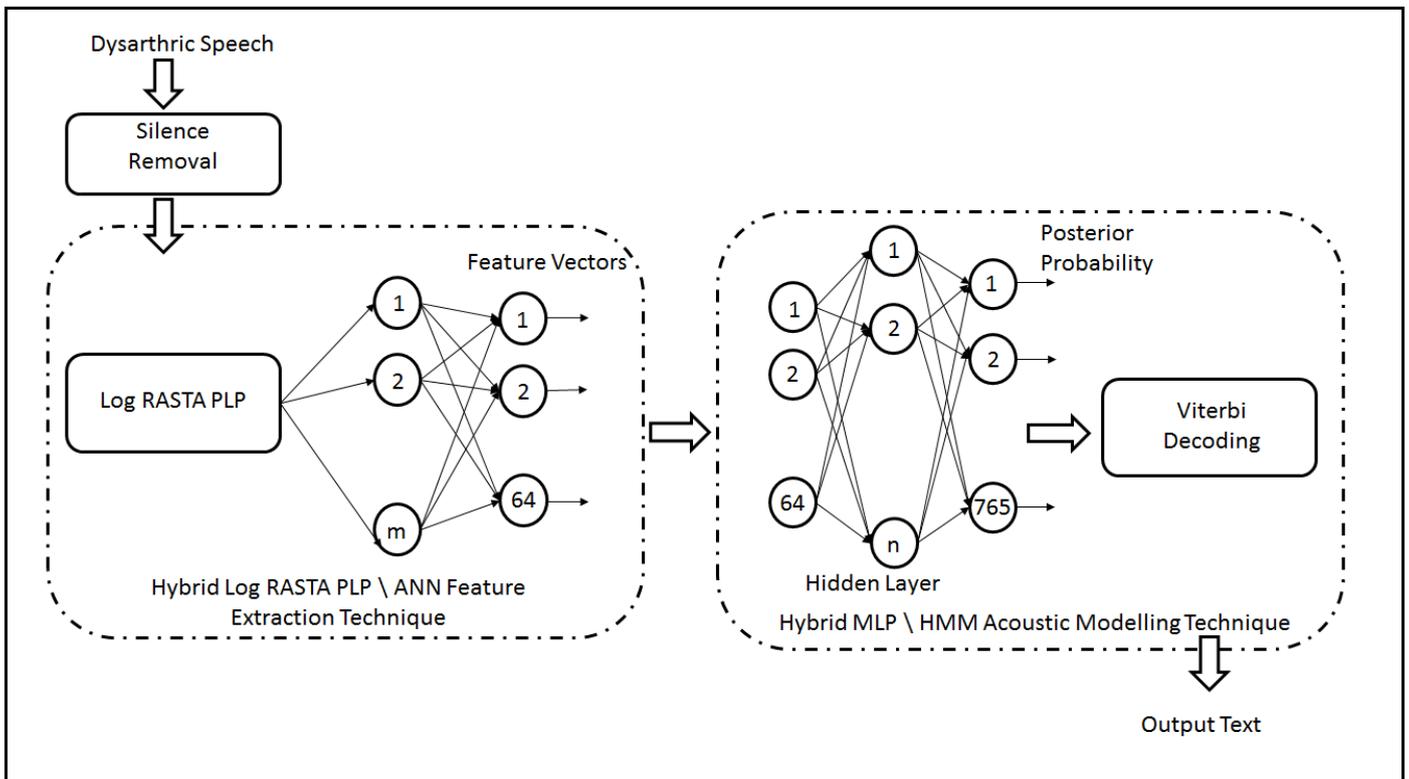

Fig.1: Hybrid Feature Extraction and Acoustic Modelling ASR for Dysarthric Speaker

*A. System Description*

For the implementation of the hybrid HMM/MLP approach acoustic feature vectors [13, 14] are need to extracted from the raw speech file. These acoustic feature vectors are extracted through feature extraction technique. After performing several experiments by using various feature extraction techniques 12-Log RASTA PLP method with frame length equal to 12 is selected for Dysarthric speech having 25ms of frame size and 10ms of overlap. Frame length is choosen to be equal to maximum length of the utterance. Silence portion is removed from the beginning and end portion based on energy of frame and frame length of each utterance is made equal by appending zeros at the end in order to make number of inputs same for each utterance to neural net. Feature extracted matrix is transferred to array form by appending m+1 column to the end of $m^{th}$ column. So, each utterance is represented by 126 features (13 features per frame x 12 frames). Each feature was assigned to one of the corresponding neuron of the clustering structure of ANN which groups features into 64 different clusters which sufficient for phoneme classification.

The clustered features are mapped to input neurons of the ANN structure for acoustic modelling. Here 3 layer MLP structure is considered with number of input neurons equals to number of features and output neurons are equal to size of the vocabulary the hidden layer consists of 5000 neurons.

The feed-forward network and backpropagation methods are used for training the features. The output of the net is converted into posterior probability for HMM modelling and applied to the Viterbi decoding along with the transition probability and prior probability. Which decodes the sequence of uttered words and output text is obtained. Detailed block diagram is shown in below Fig. 1

Above system forms the baseline system. The genetic Algorithm is applied for the optimization of the probabilities forming baseline + GA system. Mean square error is used as optimizing function and it updates probability matrix accordingly.

The training time required for the system is quiet more so, in order to reduce the time and hence the system complexity number of acoustic feature vectors are varied and tested the system performance. Variation in number of acoustic features shows distinct variation in recognition rate and system complexity. Here, system name is given based on number of feature vector considered from SOM neural net. Number of feature vector are added as suffix to the acronym of system (sys) i.e. sys16, sys32, sys64 and sys132 are having 16, 32, 64 and 128 acoustic feature vectors respectively. Sys64 is also referred to as baseline system as it was initially proposed. Application of Genetic Algorithm for optimization of HMM parameter is referred by adding suffix "+ GA" forming sys16 +GA, sys32 + GA, sys64 + GA, sys128 + GA respectively.

## B. Database Description

Database is developed at the Rehabilitation Education Centre at the University of Illinois at Urbana- Champaign. Recordings (both audio and video) took place while subjects were seated comfortably in front of a laptop computer. Subjects were asked to read an isolated word displayed on a PowerPoint slide on a computer.

The vocabulary contains 765 words including 455 distinct words and 300 distinct uncommon words chosen to maximize phone-sequence diversity. 455 distinct words contains 3 repetition of 155 words including 10 digits ("zero" to "nine"), 26 radio alphabet letters (e.g., "Alpha", "Bravo", "Charlie"), 19 computer commands (e.g., "backspace", "delete", "enter") and 100 common words (the most common words in the Brown corpus of written English such as "it", "is", "you"). The uncommon words (e.g., "naturalization", "moonshine", "exploit") were selected from children's novels digitized by Project Gutenberg, using a greedy algorithm that maximized token counts of infrequent bi-phones.

Table I summarizes the characteristics of 19 subjects that have been recorded so far. The letter M (Male) and F (Female) in speaker code specifies a participant's gender. Speech intelligibility (severity of speech disorder) is based on word transcription tasks by human listeners.

TABLE I: DATABASE DESCRIPTION

| Participant | Sex | Age | Severity of Dysarthria |
|---|---|---|---|
| M01 | Male | >18 | High |
| M04 | Male | >18 | High |
| M05 | Male | 21 | Moderate |
| M06 | Male | 18 | Moderate |
| M07 | Male | 58 | High |
| M08 | Male | 28 | Mild |
| M09 | Male | 18 | Mild |
| M10 | Male | 21 | Mild |
| M11 | Male | 48 | Moderate |
| M12 | Male | 19 | High |
| M14 | Male | 44 | Mild |
| M16 | Male | 40 | High |
| F02 | Female | 30 | High |
| F03 | Female | 51 | High |
| F04 | Female | 18 | Moderate |
| F05 | Female | 22 | Mild |

## IV. RESULT & DISCUSSION

In this section we describe the result obtained on performing the various experiment and detailed discussion is made on it. Here performance evaluation of proposed four system is carried out.

The evaluation criteria considered for testing system performance is Word Recognition Rate (WRR) or Word Recognition Accuracy (WRA) which describes the correctness of system performance from speaker point of view. It is calculated as below:

$$WRA(\%) = \left(\frac{W_{TOT} - W_{err}}{W_{TOT}}\right) \times 100$$

Where,

$W_{TOT}$ = Total number of words
$W_{err}$ = Number if incorrect recognized words

The system described in section III is having 64 feature vectors obtained from SOM Neural Net which act as an input to MLP, but the training time required for this system is fairly high. So, in order to decrease the complexity of the system number of feature vectors are changed, maintaining other parameters and its performance is evaluated. Here, 16, 32 and 128 features vectors are tested forming sys16, sys32 and sys128 respectively. System with 64 feature is either known as Baseline system or sys64.

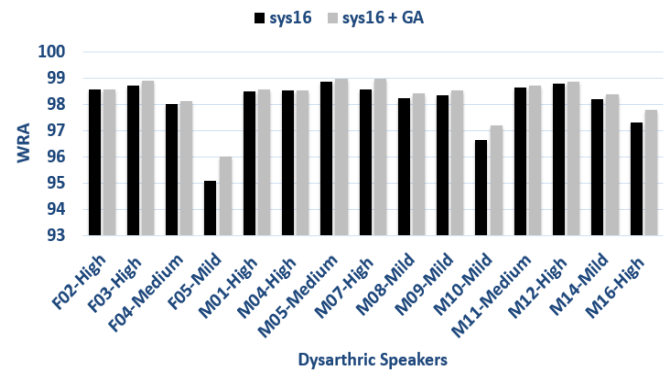

Fig.2: WRA comparison of sys16 and sys16 + GA

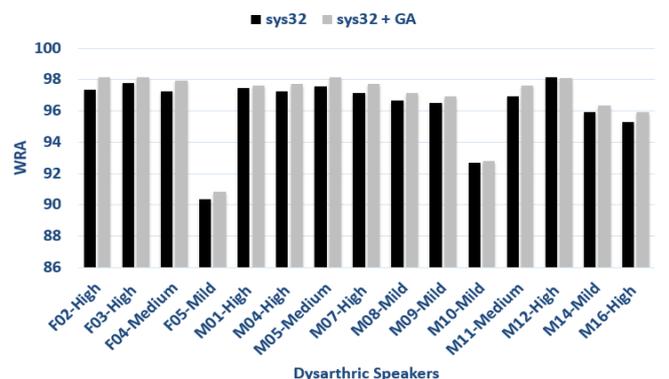

Fig.3: WRA comparison of sys32 and sys32 + GA

From Fig. 2 to Fig. 5 shows the comparison of WRA for baseline and baseline optimized with Genetic Algorithm (GA) i.e. baseline + GA. Here, patients along with severity in Dysarthria is represented is represented in X-axis while recognition rate is represented at Y-axis. It can be observed that WRA differs among the speakers with F05 having least recognition rate. Also, optimization of posterior probability through GA improves the performance. This is observed among all adopted system but the range of recognition rate differs from each other. For all four system F05 and M10 speakers are found to get least trained and have comparatively less recognition rate in reference to other speakers but significantly higher in order to practice this system on daily basis.

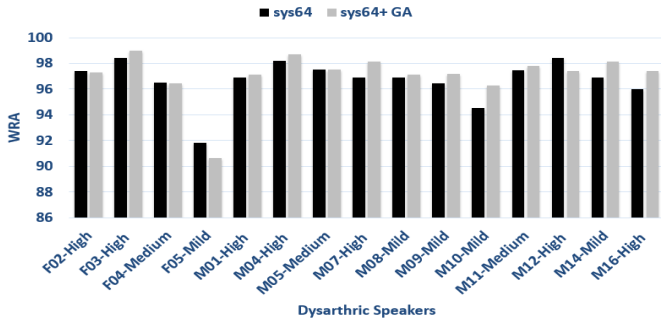
Fig.4: WRA comparison of sys64 and sys64 + GA

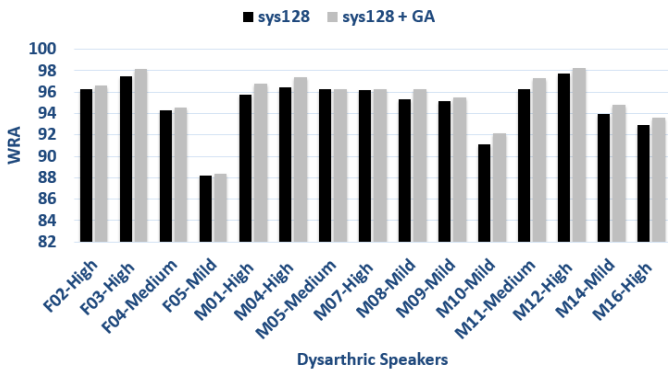
Fig.5: WRA comparison of sys128 and sys128 + GA

Below Fig. 6 shows the overall performance of all the system along with evolution of observation probabilities through GA. It is found experimentally that system with less number of feature vectors gives highest recognition rate and vice versa. This is due to constant number of hidden and output nodes in feed forward neural network and varying number of input acoustic features. Lower number of input feature i.e. less number of input nodes is trained with higher hidden and output nodes makes the system to perform better as compared to others.

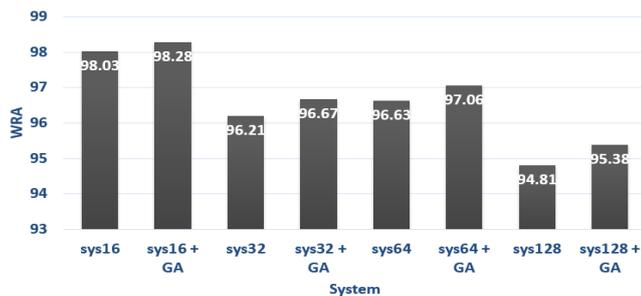
Fig.6: WRA Comparison of all System

Table II shows the comparison of different proposed system in terms of system complexity i.e. time required for training neural network along with obtained recognition rate. It is observed that as the number of feature vector increases training time for SOM network of clustering increases as there are more number of groups to get into classified while opposite is the case for MLP network due to decrease in difference between number of input and hidden nodes but, overall training period tends to depend more on training of feed forward network as it takes much training period as compared to SOM topology. Observation from the table indicates that recognition rate and training period is almost same for sys32 and sys64. Detailed description of each system along with speaker is summarized in tabular format attached in appendix

TABLE II: WRA & TRAINING TIME COMPARISON

| System | Required Time (Hr:Min:Sec) | | Total Time Required (Hr:Min:Sec) | WRA (%) |
|---|---|---|---|---|
| | SOM Network | Feed Forward Network | | |
| sys16 | 8:02 | 5:22:15 | 5:30:17 | 98.03 |
| sys32 | 14:47 | 3:04:14 | 3:19:01 | 96.21 |
| sys64 | 30:53 | 2:51:12 | 3:22:05 | 96.63 |
| sys128 | 1:01:13 | 2:43:46 | 3:44:51 | 94.81 |

CONCLUSION

In this paper we studied that hybridized approach for feature extraction and acoustic modelling technique is found to increase the system performance. System performance is further boosted by the optimization of the parameters through genetic algorithm. Also it is observed that there is less variation in WRA among speakers having different severity in dysarthria. Number of feature vectors plays an important role for system complexity and WRA of the system. As the number of feature vectors increases, WRA decreases and total training period which is the combination of time required for SOM and MLP network also decreases. Also, it is observed that performance for system having 32 and 64 acoustic features remains almost constant having accuracy 96.21% and 96.63% of WRA respectively. Highest WRA is obtained for system having 16 feature vectors optimized with genetic algorithm i.e. 98.28% with highest consumed training time of 5:30:17.

APPENDIX

Tables shown below summarized performance of all the described system for each speaker. TABLE shows the performance for basic four system i.e. sys16, sys32, sys64 and sys128. Total 765 unique words are used and each word is repeated about 5-7 times forming different number of total words for which system is trained. Number of words which are falsely classified are mentioned in 4$^{th}$ column of both the table and last column indicates WRA for each system. TABLE shows the same description of all four system whose posterior probability is optimized along with Genetic Algorithm forming sys-x + GA, where x = 16,32,64,128.

TABLE I: DETAILED PERFORMANCE ANALYSIS OF SYSTEM

| Speakers | Severity of Dysarthria | Total number of Words | Number of Incorrectly Classified Words | | | Word Recognition rate (%) | | |
|---|---|---|---|---|---|---|---|---|
| | | | sys16 | sys32 | sys128 | sys16 | sys32 | sys128 |
| F02 | High | 5355 | 77 | 142 | 199 | 98.56 | 97.35 | 96.28 |
| F03 | High | 5355 | 69 | 119 | 137 | 98.71 | 97.78 | 97.44 |
| F04 | Moderate | 5355 | 106 | 148 | 306 | 98.02 | 97.24 | 94.29 |
| F05 | Mild | 5348 | 262 | 515 | 629 | 95.1 | 90.37 | 88.24 |
| M01 | High | 2805 | 43 | 71 | 118 | 98.47 | 97.47 | 95.79 |
| M04 | High | 3825 | 56 | 106 | 135 | 98.54 | 97.23 | 96.47 |
| M05 | Moderate | 5355 | 61 | 129 | 202 | 98.86 | 97.59 | 96.23 |
| M07 | High | 5355 | 77 | 152 | 203 | 98.56 | 97.16 | 96.21 |
| M08 | Mild | 5355 | 94 | 177 | 252 | 98.24 | 96.69 | 95.29 |
| M09 | Mild | 5355 | 89 | 188 | 258 | 98.34 | 96.49 | 95.18 |
| M10 | Mild | 5354 | 181 | 392 | 475 | 96.62 | 92.68 | 91.13 |
| M11 | Moderate | 4590 | 62 | 141 | 172 | 98.65 | 96.93 | 96.25 |
| M12 | High | 4590 | 56 | 86 | 106 | 98.78 | 98.13 | 97.69 |
| M14 | Mild | 5355 | 97 | 219 | 322 | 98.19 | 95.91 | 93.99 |
| M16 | High | 4590 | 123 | 217 | 324 | 97.32 | 95.27 | 92.94 |
| | | | | | | | | |
| Total | | 73942 | 1453 | 2802 | 3838 | 98.03 | 96.21 | 94.81 |

TABLE II: DETAILED PERFORMANCE ANALYSIS OF GA OPTIMIZED SYSTEM

| Speakers | Severity of Dysarthria | Total Number of Words | Number of Incorrectly Classified Words | | | Word Recognition Rate (%) | | |
|---|---|---|---|---|---|---|---|---|
| | | | sys16+GA | sys32+GA | sys128+GA | sys16+GA | sys32+GA | sys128+GA |
| F02 | High | 5355 | 77 | 100 | 180 | 98.56 | 98.13 | 96.64 |
| F03 | High | 5355 | 60 | 100 | 100 | 98.88 | 98.13 | 98.13 |
| F04 | Moderate | 5355 | 100 | 110 | 290 | 98.13 | 97.95 | 94.58 |
| F05 | Mild | 5348 | 213 | 490 | 620 | 96.02 | 90.84 | 88.41 |
| M01 | High | 2805 | 40 | 67 | 90 | 98.57 | 97.61 | 96.79 |
| M04 | High | 3825 | 56 | 87 | 100 | 98.54 | 97.73 | 97.39 |
| M05 | Moderate | 5355 | 56 | 99 | 200 | 98.95 | 98.15 | 96.27 |
| M07 | High | 5355 | 55 | 123 | 200 | 98.97 | 97.70 | 96.27 |
| M08 | Mild | 5355 | 85 | 154 | 200 | 98.41 | 97.12 | 96.27 |
| M09 | Mild | 5355 | 80 | 164 | 240 | 98.51 | 96.94 | 95.52 |
| M10 | Mild | 5354 | 150 | 385 | 420 | 97.20 | 92.81 | 92.16 |
| M11 | Moderate | 4590 | 60 | 110 | 125 | 98.69 | 97.60 | 97.28 |
| M12 | High | 4590 | 53 | 88 | 79 | 98.85 | 98.08 | 98.28 |
| M14 | Mild | 5355 | 88 | 196 | 280 | 98.36 | 96.34 | 94.77 |
| M16 | High | 4590 | 102 | 187 | 295 | 97.78 | 95.93 | 93.57 |
| | | | | | | | | |
| Total | | 73942 | 1275 | 2460 | 3419 | 98.28 | 96.67 | 95.38 |